\documentclass[copyright,creativecommons]{eptcs}
\providecommand{\event}{WWV 2011} 

\usepackage[pdftex]{graphicx}

\usepackage{amssymb}
\usepackage{amsmath}%
\usepackage{amsfonts}%
\usepackage{cmll} 
\usepackage{verbatim}
\usepackage{setspace}
\usepackage{proof}
\usepackage{array}
\usepackage{changebar}

\DeclareMathOperator{\plusL}{L}
\DeclareMathOperator{\plusR}{R}

\DeclareMathOperator{\cc}{::\ }
\DeclareMathOperator{\cln}{:}

\title{A theorem proving framework for the formal verification of Web Services Composition}
\author{Petros Papapanagiotou \qquad\qquad Jacques D. Fleuriot
\institute{School of Informatics\\
University of Edinburgh \\
Informatics Forum, 10 Crichton Street\\
Edinburgh EH8 9AB, UK}
\email{P.Papapanagiotou@sms.ed.ac.uk \quad\qquad jdf@inf.ed.ac.uk}
}
\def\titlerunning{A theorem proving framework for the formal verification of web services composition}
\def\authorrunning{P. Papapanagiotou, J.D. Fleuriot}
\begin{document}
\maketitle

\begin{abstract}
We present a rigorous framework for the composition of Web Services within a higher order logic theorem prover. Our approach is based on the proofs-as-processes paradigm that enables inference rules of Classical Linear Logic (CLL) to be translated into $\pi$-calculus processes. In this setting, composition is achieved by representing available web services as CLL sentences, proving the requested composite service as a conjecture, and then extracting the constructed $\pi$-calculus term from the proof. Our framework, implemented in HOL Light, not only uses an expressive logic that allows us to incorporate multiple Web Services properties in the composition process, but also provides guarantees of soundness and correctness for the composition. 
\end{abstract}

\section{Introduction}
\label{intro}

The general aim of the current research is to design and implement a rigorous framework for the composition and formal verification of Web Services based on higher order logic. Our approach is motivated by recent work on the automated composition of Semantic Web Services using Intuitionistic Linear Logic that has shown promising results \cite{rao-semantic,rao2006composition}.

We focus mainly on the complex task of quality-driven Web Services composition. This involves the appropriate collection and combination of multiple Web Services in order to achieve a composite service that can perform a complex task. The composition needs to take into consideration non-functional restrictions, including location, cost, and time, and be quality-driven because the system should ensure a user-specified Quality of Service based on the quality provided in each of the participating Web Services descriptions. The complexity of the task is compounded by the dramatic increase in available Web Services, as well as the great variety of conceptual models used for the descriptions of the services. 

\subsection{Overview}
\label{overview}

Our aim is to deploy our system as an automated, offline (as opposed to on-the-fly) Web services composer. Using an expressive logic allows the system to incorporate all of the aforementioned information when composing the services. However, this also means a need to maintain a balance in the tradeoff between expressiveness and decidability. We believe the latter to be important since decidability directly affects the degree of automation and, therefore, the user-friendliness of the system. 

The compositions are accomplished using the proofs-as-processes paradigm as introduced by Abramsky \cite{abramsky1994proofs} and Bellin and Scott \cite{bellin1994}. Abramsky showed the relevance between a Classical Linear Logic (CLL) \cite{girard1995linear} proof and the $\pi$-calculus \cite{milner1999communicating} by modifying the Curry-Howard isomorphism and using the formulae-as-types paradigm \cite{howard1980formulae}. Proofs are viewed as $\pi$-calculus processes (instead of $\lambda$-calculus functions). Bellin and Scott formalised Abramsky's translation of a fragment of CLL to $\pi$-calculus and provided proofs of soundness and correctness of the translation. 

We exploit this translation by producing Web Services compositions as CLL proofs, with the requested service set as a conjecture in each case. The $\pi$-calculus representation of the composite service is then extracted by translating the proof based on the proofs-as-processes paradigm.

Our implementation is being developed within the higher order proof assistant HOL Light \cite{harrison1996hol}. The system has equality as the only primitive concept and a few primitive inference rules that form the basis of more complex rules and tactics. Built on top of these, HOL Light has automated methods for proofs such as model elimination \cite{harrison1996optimizing} and decision procedures. Additionally, it has an array of \emph{conversion} methods that allow for very efficient and fine-grained manipulation (such as rewriting or numerical reduction) and automatic proofs of formulas. The system is based on the LCF approach \cite{paulson1990logic}, which, guarantees that any proved theorem is a logical consequence of the primitive axioms.

\subsection{Motivating Example}

\begin{figure}[htbp]
	\centering
		\includegraphics[scale=0.4]{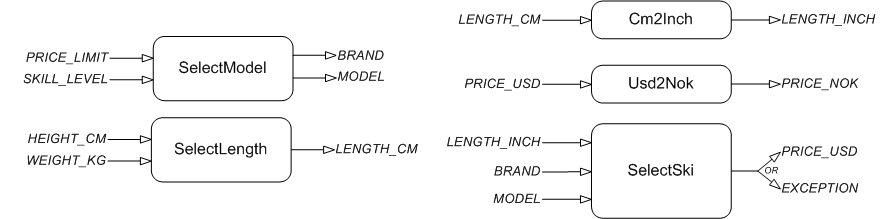}
	\caption{Diagrams of the available services for the Ski example.}
	\label{fig:AvailServ}
\end{figure}

\begin{figure}[htbp]
	\centering
		\includegraphics[scale=0.4]{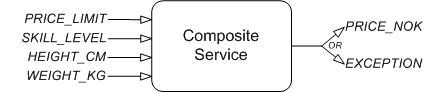}
	\caption{Diagram of the requested service for the Ski example.}
	\label{fig:ReqServ}
\end{figure}

We consider the case of ordering a ski set, as presented by Rao et al.\ \cite{rao2006composition}. In this, we compose a core service with value-adding services, ie.\ services that have minor, independent functionality, such as currency or measurement conversion, that can be used as addons to the core functionality. The core service ``selectSki'', returns the price in US dollars of a ski set using the ski length, brand, and model as input parameters. There are also various value-adding services whose functionality is demonstrated in the diagram provided in Figure \ref{fig:AvailServ}. 
The requested composite service must return the price of a ski set in Norwegian Crowns (NOK) given the user's height, weight, and skill level as well as a price limit. The diagram of the requested service can be found in Figure \ref{fig:ReqServ}.
Some non-functional attributes have been attached to the services. A cost of 10 NOK is attached to ``selectBrand'' and 20 NOK for ``selectModel'', the ``USD2NOK'' service only replies to requests that are certified by Microsoft and the ``selectSki'' service is located in Norway (note that this information is not apparent in the diagrams).

In Section \ref{background}, we will describe the theoretical background of our work by briefly explaining the main concepts of CLL and the $\pi$-calculus. Then, in Section \ref{paradigm}, we introduce the CLL to $\pi$-calculus translations of the original proofs-as-processes paradigm and give an intuitive interpretation for some of the rules. In Section \ref{results} we present the results we obtained from our system for the Ski example, followed by an overview of the related work in Section \ref{relatedwork}. We conclude in Section \ref{conclusion} with our plans for future work.

\section{Background}
\label{background}

In this section, we discuss some of the theoretical background of the proofs-as-processes paradigm. In particular, we briefly explain the main concepts of Classical Linear Logic in Section \ref{CLL} and $\pi$-calculus in Section \ref{pi}. We focus on how these two languages can be used to describe Web Services, which is the first step towards achieving Web services composition using the proofs-as-processes paradigm. 

\subsection{Classical Linear Logic}
\label{CLL}

Girard proposed linear logic (LL) as a refinement to classical logic \cite{girard1995linear}. In LL, the emphasis is not merely on the truth of a statement as in the classical logic but also on formulas that represent resources. The classical rules of contraction and weakening are not allowed in LL and therefore assumptions cannot be ignored or copied. For example, if a constant $A$ is assumed twice, it is considered a distinct case than when it is assumed once. In order to achieve a proof, all assumptions must be ``consumed'' as resources. Contraction and weakening rules are only used on assumptions with additional modal connectives called exponentials, such as the ``of-course'' operator ``!''. The use of these connectives allows for classical logic to be encoded within LL. In computer science, LL has been used as a direct and declarative approach to reasoning about various computational models related to services such as Petri Nets \cite{murata1989petri}.

\subsubsection{Description}
\label{CLLdesc} 

For the purposes of Web Services representation and composition, we aim to use propositional Classical Linear Logic (CLL). This version of LL includes multiplicative conjunction and disjunction, additive conjunction and disjunction, linear negation and the of-course and why-not operators (also referred to as exponentials). All these operators can be intuitively interpreted in the context of resources for Web services and we informally discuss the semantics for some of them next:

\begin{itemize}

\item Multiplicative conjunction or the ``tensor'' operator ($A \otimes B$) indicates a simultaneous operation which, in the context of resources, refers to the simultaneous production of $A$ and $B$. In order to prove $A \otimes B$, the set of available resources must be split in two subsets, one that can achieve $A$ and one that can achieve $B$. Multiplicative conjunction can be seen as the counterpart of conjunction in classical logic. In the context of Web Services, multiplicative conjunction can also be used to represent quantities of consumable resources such (most typically) money and time. 

In our Ski example, the \texttt{selectModel} service outputs the selected brand and a particular model simultaneously. If we represent these outputs as resources in CLL, the \texttt{selectModel} service output would be $(BRAND\otimes MODEL)$.

\item Additive disjunction or the ``plus'' operator ($A \oplus B$) can be viewed as the equivalent of exclusive disjunction in classical logic and indicates that either of $A$ or $B$ are produced but not both. When representing Web Services, additive disjunction can be used to indicate alternative results. Most typically it is used to express the possibility of a Web Service throwing an exception instead of producing the expected result. It is worth mentioning that most web services composition methodologies do not take exceptions into consideration.

In our example, the \texttt{selectSki} service may fail to return the price of the selected ski if, for example, the particular model is out of stock or is not available for the given length. In this case, \texttt{selectSki} will output an exception. In CLL we can represent the output of \texttt{selectSki} as $(PRICE\_USD\oplus EXCEPTION)$.

\item Linear negation ($\cdot ^\bot$) in CLL obeys similar laws to those of classical negation. The symmetry of CLL becomes evident in the connection of the dual operators through linear negation. For example, negating the tensor operator ($(A \otimes B)^\bot$) results in a ``par'' operator ($A^\bot \parr B^\bot$).

In general, we use negated CLL terms to represent input (as opposed to output for non-negated terms). For example, the input of the \texttt{Cm2Inch} process can be represented as $(LENGTH\_CM^\bot)$.

\item The of-course ($!A$) and why-not ($?A$) operators (also refered to as ``exponentials'') are used to represent unlimited resources. They allow controlled versions of the weakening and contraction rules. Essentially, these correspond to the replication of a resource as many times as necessary. For example, in the context of Web Services, functional parameters such as input and output are reusable in contrast to states which, once ``consumed'' through a state transition, are no longer applicable. 
\end{itemize}

Generally, a two-sided sequent calculus is used for the representation of the CLL inference rules. The left and right versions of each inference rule serve the purpose of handling a connective on the left or right hand side of the turnstyle respectively. However, given the observation that $\Gamma \vdash \Delta$ is equivalent to $\vdash \Gamma^\bot,\ \Delta$ we can eliminate half of the rules by using a one-sided sequent calculus representation. This has an important impact in the automation of CLL proofs, as the number of available inference rules in the proof search is effectively halved. We note that Bellin and Scott also use a one-sided sequent calculus representation for CLL in their work.

The one-sided sequent calculus versions of the inference rules for the multiplicative-additive fragment (MALL) of CLL (ie.\ without the exponentials) are presented in Figure \ref{CLLpi}. We note that, in this particular figure, the rules are annotated using process calculus channel names (see Section \ref{paradigm} for more details). From here on in this paper, unless otherwise stated, every reference to CLL corresponds to MALL as this is the fragment we are currently focused on. In the future, we plan to increase the expresiveness of our system by extending the logic to the full CLL, although the latter is undecidable \cite{john1992decision}. 

\subsubsection{Describing Web Services using CLL}
\label{WSCLL}

Inspired by the translation of Web Services to Intuitionistic Linear Logic (ILL) proposed by Rao \cite{rao-semantic}, we utilise a similar approach for CLL. The CLL syntax has significant differences from ILL though; the most important being the lack of linear implication. As already mentioned, we have chosen to interpret negated terms as input and normal terms as output since this appears somewhat more intuitive\footnote{The duality of the CLL connectives allows this arbitrary choice \cite{bellin1994}.}. The resulting formula is shown in Figure \ref{fig:CLLSWS}. Note that we have only kept functional properties of web services, ie.\ input ($\vec{I}$), output ($\vec{O}$), preconditions ($\vec{P}$) and effects ($\vec{F}$). This formula can be expanded to incorporate non-functional properties, such as cost and time.

\begin{figure}[htbp]
	\centering
		\[ \vdash\ \vec{P}, \vec{I}, ((\underset{i}{\otimes}\vec{F})\otimes (\underset{i}{\otimes}\vec{O})) \oplus E) \]
		Where: $\underset{i}{\otimes} (a_1,a_2,..., a_n) = a_{1} \otimes a_{2} \otimes ... \otimes a_{n}$
	\caption{The CLL representation of a Web Service.}
	\label{fig:CLLSWS}
\end{figure} 

Following this formula, the translation of the available services and the request service of the Ski example (see Figures \ref{fig:AvailServ} and \ref{fig:ReqServ}) in CLL are shown in Figure \ref{fig:AvailServCLL}. We note that these translations are annotated with process calculus terms using the ``$\cc$'' and ``$\cln$'' operators. These are explained in Section \ref{papdesc}. 

\begin{figure}[htbp]
\begin{center}
		Available services: \\
		\medskip
\begin{minipage}{.75\linewidth}
\scriptsize
	$\vdash SelectModel\cc smp \cln PRICE\_LIMIT^\bot,\ sms \cln SKILL\_LEVEL^\bot,\ sso \cln (BRAND\otimes MODEL)$ \\
	$\vdash SelectLength \cc slh \cln HEIGHT\_CM^\bot,\ slw \cln WEIGHT\_KG^\bot,\ sll \cln LENGTH\_CM$ \\
	$\vdash Cm2Inch \cc cic \cln LENGTH\_CM^\bot,\ cii \cln LENGTH\_IN$ \\
	$\vdash Usd2Nok \cc unu \cln PRICE\_USD^\bot,\ unn \cln PRICE\_NOK$ \\
	$\vdash SelectSki \cc ssl \cln LENGTH\_IN^\bot,\ ssb \cln BRAND^\bot,\ ssm \cln MODEL^\bot,\ sso \cln (PRICE\_USD\oplus EXCEPTION)$ \\
\end{minipage}
\end{center}
\normalsize	  
\begin{center}
Request:  \\
\medskip
\begin{minipage}{.9\linewidth}
\scriptsize
		$\vdash P \cc x \cln PRICE\_LIMIT^\bot,\ y \cln SKILL\_LEVEL^\bot,\ z \cln HEIGHT\_CM^\bot,\ w \cln WEIGHT\_KG^\bot,\ t \cln (PRICE\_NOK \oplus EXCEPTION)$ \\
\end{minipage}
\end{center}
\normalsize
	\caption{The available services and the request for the Ski example translated into CLL with process annotations.}
	\label{fig:AvailServCLL}
\end{figure}

This formula also forms the basis for the translation of Web Services described in a variety of languages, including WSDL \cite{christensen2001web}, BPEL4WS \cite{andrews2003business} or, in the case of Semantic Web Services, OWL-S \cite{martin2004owl}, the follow-up of DAML-S \cite{coalition2002daml} (used by Rao et al.\ as part of their work), into CLL. The details of these translations are beyond the scope of the current paper though.

\subsection{The $\pi$-calculus}
\label{pi}

The $\pi$-calculus is a formalism aimed at the description of concurrent processes \cite{milner1999communicating}. The name is used to show the connection to the $\lambda$-calculus as a minimal, abstract representation. In the $\pi$-calculus, processes are described atomically as independent entities. They are attached to channels that use variables to denote the input or output of the process. 

\subsubsection{Description}
\label{pidesc}

The syntax of the polyadic $\pi$-calculus is presented by the following grammar:
\[ 
P ::= x(\vec{y}).P \,\,|\,\, \overline{x} \langle \vec{y} \rangle.P \,\,|\,\, P || P \,\,|\,\, P + P \,\,|\,\,(\nu \vec{x})P \,\,|\,\, !P \,\,|\,\, 0 
  \]
A channel $x$ attached to process $P$ that allows it to receive a message that will be bound to the vector of names $\vec{y}$ is represented as $x(\vec{y}).P$. Similarly, $\overline{x} \langle \vec{y} \rangle.P$ depicts a process with channel $x$ that can send a message through name $\vec{y}$ as output. It is worth noting that the infix dots in these two cases are often omitted for simplicity. The expression $P || P$ describes the parallel composition of two processes whereas $(\nu \vec{x})P$ describes a vector of names $\vec{x}$ that is local to $P$. Finally, $!P$ describes a process $P$ that can be replicated and $0$ represents the ``$nil$'' process that has no functionality. Note that in some more minimal versions, the non-deterministic choice $P\ +\ Q$ between two processes $P$ and $Q$ is excluded from the syntax. It is worth remarking also, that there is no explicit representation of sequential processes. Interactions between parallel processes are represented as reductions of the $\pi$-calculus terms (similar in form to the reductions of $\lambda$-calculus). There are other extensions to the $\pi$-calculus syntax and other important concepts such as bisimulation equivalence (or bisimilarity) \cite{sangiorgi1996theory} which go beyond the scope of the current work.

Reductions of $\pi$-calculus terms are defined formally using a set of rules. We will only present the most significant reduction rule that describes the interaction between two parallel processes:
\begin{equation}
\label{pireduction}
 (... + x(\vec{a}).F) || (... + \overline{x} \langle \vec{b} \rangle .C) \rightarrow F[\vec{b}/\vec{a}] || C
\end{equation}

The communication described here happens between process $C$ with output $\vec{b}$ over channel $x$ and process $F$ with input $\vec{a}$ over the same channel $x$. The processes run in parallel (denoted by the ``$||$'' symbol) and for their interaction $C$ sends $\vec{b}$ to $F$ over $x$ yielding $F||C$ where each free occurence of the names in $\vec{a}$ in $F$ is replaced by the names in $\vec{b}$. It must be noted that the above interaction is only allowed if the two involved vectors $\vec{a}$ and $\vec{b}$ have the same size.

Moreover, in addition to the reduction rule, a set of rules is defined that allow us to express structural congruence ($\equiv$) relations between processes. For example, $P+0\equiv 0+P\equiv P$, $P||0\equiv 0||P\equiv P$ and $!P\equiv P||!P$ are all defined as structural congruence rules. This concludes a general overview of the $\pi$-calculus, its reductions, and its intuitive interpretation.

The $\pi$-calculus has formed the basis for a variety of process algebras used to describe the communication between agents, including LCC \cite{robertson2004lightweight} and BPEL4WS \cite{andrews2003business}. Additionally, there are multiple available tools that perform a variety of tasks involving $\pi$-calculus terms. For example, the Mobility Workbench (MWB) \cite{victor1994mobility} performs checks for open bisimulation equivalences \cite{sangiorgi1996theory} (which roughly corresponds to checking agents for equivalent behaviour) and the PiVizTool \cite{bog2006tool} is a tool written in Java that can graphically represent agents described in $\pi$-calculus and allows a step by step, user-controlled monitoring of the interactions in a multi-agent environment.

\section{The proofs-as-processes paradigm}
\label{paradigm}

In the proofs-as-processes paradigm, Bellin and Scott \cite{bellin1994} give a corresponding $\pi$-calculus term for each of the CLL inference rules. As the inference rules are applied within a proof, these correspondences allow the construction of a $\pi$-calculus term that corresponds to the entire proof. At the end of the proof, it is guaranteed that applying the possible reductions at the resulting $\pi$-calculus term corresponds to the process of cut-elimination in the proof. This means that the cut-free version of the proof corresponds to an equivalent $\pi$-calculus term that cannot be reduced further.

\subsection{Description}
\label{papdesc}

Bellin and Scott attach free variables as proof annotations to every CLL sequent. Each of these variables corresponds to a $\pi$-calculus communication port. Moreover, the process calculus term attached to the rule is dependent on the processes attached to each of the premises of the rule and the process annotations of the involved sequents. For example, let us consider the ``tensor'' rule for CLL including the process annotations as shown below:

\begin{equation}
		\infer[\otimes]{\vdash \vec{w} \cln \Gamma,\ \vec{u} \cln \Delta,\ z \cln A\otimes B}{
		\infer*[F]{\vdash \vec{w} \cln \Gamma,\ x \cln A}{}
		&
		\infer*[G]{\vdash \vec{u} \cln \Delta,\ y \cln B}{}
		}
\end{equation}
Processes $F$ and $G$ are attached to the two premises of the rule based on any previous proof steps. The process calculus term attached to the rule, also refered as the ``translation'' of the rule to the $\pi$-calculus is given by the following term:
\begin{equation}
\label{tensortrans}
\overset{x,y}{\underset{z}{\bigotimes}}(F,G)\vec{w}\vec{u}z \equiv \nu x y (\overline{z}\langle xy \rangle(F_{\vec{w}x} || G_{\vec{u}y}))
\end{equation}

We note that term (\ref{tensortrans}) is dependent on processes $F$ and $G$ and also on the channel names $x$, $y$, $\vec{w}$ and $\vec{u}$ attached to the involved sequents. The free variables found in the annotations of the sequents of a conclusion of an inference rule are ensured to be exactly the same as the free names of the corresponding $\pi$-calculus term for that rule, in this case $z$, $\vec{w}$, and $\vec{u}$. Following a syntax closer to the one used in type theory, we can also represent the same annotated rule and corresponding translation using the $\cc$ operator as follows:

\begin{equation}
		\infer[\otimes]{\vdash \overset{x,y}{\underset{z}{\bigotimes}}(F,G)\vec{w}\vec{u}z \cc \vec{w} \cln \Gamma,\ \vec{u} \cln \Delta,\ z \cln A\otimes B}{
		\vdash F \cc \vec{w} \cln \Gamma,\ x \cln A
		&
		\vdash G \cc \vec{u} \cln \Delta,\ y \cln B
		}
\end{equation}

The seven basic CLL inference rules and their process correspondences are shown in Figure \ref{CLLpi}. Before analyzing and giving a practical, intuitive explanation for some of these correspondences, we should note the two choices made in them. The symmetry of CLL allows for two equivalent, symmetric $\pi$-calculus translations for the identity axiom and the $\otimes$ and $\parr$ operators. We have chosen to translate positive atoms and the $\otimes$ operator as senders while we translate negative atoms and the $\parr$ operator as receivers. We examine some of these rules from Figure \ref{CLLpi} and their translations more closely next:

\begin{figure}[htb]
	\centering
	\footnotesize
		\begin{tabular}{>{\centering\arraybackslash}m{.4\linewidth} >{\centering\arraybackslash}m{.55\linewidth}}
		CLL inference rule & $\pi$-calculus translation \\
		\vspace*{3mm}
		$\vdash x \cln A,\ y \cln A^\bot$ & $Ixy\equiv y(a)\overline{x}\langle a \rangle$ \\
		\vspace*{-5mm}
		$$
		\infer[\otimes]{\vdash \vec{w} \cln \Gamma,\ \vec{u} \cln \Delta,\ z \cln A\otimes B}{
		\infer*[F]{\vdash \vec{w} \cln \Gamma,\ x \cln A}{}
		&
		\infer*[G]{\vdash \vec{u} \cln \Delta,\ y \cln B}{}
		}
		$$ & 
		$\overset{x,y}{\underset{z}{\bigotimes}}(F,G)\vec{w}\vec{u}z \equiv \nu x y (\overline{z}\langle xy \rangle(F_{\vec{w}x} || G_{\vec{u}y}))$ \\
		\vspace*{-7mm}
		$$
		\infer[\parr]{\vdash \vec{w} \cln \Gamma,\ z \cln A \parr B}{
		\infer*[F]{\vdash \vec{w} \cln \Gamma,\ x \cln A,\ y \cln B}{}
		}
		$$ & $\overset{x,y}{\underset{z}{\bigparr}}(F)\vec{w}z \equiv z(xy)F_{\vec{w}xy}$ \\
		\vspace*{-7mm}
		$$
		\infer[\oplus]{\vdash \vec{w} \cln \Gamma,\ z \cln A \oplus B}{
		\infer*[P]{\vdash \vec{w} \cln \Gamma,\ x \cln A}{}
		}
		$$ & $\overset{x}{\underset{z}{\plusL}}(P)\vec{w}z \equiv \nu x (z(uv)\overline{u}\langle x \rangle P_{\vec{w}x})$ \\
		\vspace*{-7mm}
		$$
		\infer[\oplus]{\vdash \vec{w} \cln \Gamma,\ z \cln A \oplus B}{
		\infer*[Q]{\vdash \vec{w} \cln \Gamma,\ y \cln B}{}
		}
		$$ & $\overset{y}{\underset{z}{\plusR}}(Q)\vec{w}y \equiv \nu y (z(uv)\overline{v}\langle y \rangle Q_{\vec{w}y})$ \\
		\vspace*{-7mm}
		$$
		\infer[\with]{\vdash \vec{w} \cln \Gamma,\ z \cln A\with B}{
		\infer*[P]{\vdash \vec{w} \cln \Gamma,\ x \cln A}{}
		&
		\infer*[Q]{\vdash \vec{w} \cln \Gamma,\ y \cln B}{}
		}
		$$ & 
		$\overset{x,y}{\underset{z}{\bigwith}}(P,Q)\vec{w}z \equiv \nu u v (\overline{z}\langle uv \rangle[u(x)P_{\vec{w}x} + v(y)Q_{\vec{w}y}])$ \\
		\vspace*{-7mm}
		$$
		\infer[Cut]{\vdash \vec{u} \cln \Gamma,\ \vec{v} \cln \Delta}{
		\infer*[F]{\vdash \vec{u} \cln \Gamma,\ x \cln C}{}
		&
		\infer*[G]{\vdash \vec{v} \cln \Delta,\ y \cln C^\bot}{}
		}
		$$ & 
		$Cut^{z}(F,G)\vec{u}\vec{v} \equiv \nu z (F_{\vec{u}}[z/x] || G_{\vec{v}}[z/y])$ \\
		\end{tabular}
			\caption{The CLL inference rules annotated with channel names and the corresponding $\pi$-calculus processes.}
			\label{CLLpi}
\end{figure}

\paragraph{The identity axiom} The identity axiom $\vdash x \cln A,\ y \cln A^\bot$ can be intuitively translated given the aforementioned choices to $y(a)\overline{x}\langle a \rangle$. The resulting process receives a message $a$ through the channel $y$ of the negative literal and sends the \textit{same} message $a$ through the channel of the positive atom $x$. Such a process is refered to as an \textit{axiom buffer}.

\paragraph{The $\otimes$ rule} The ``tensor'' rule must intuitively correspond to a channel $z$ that sends two messages $x$ and $y$ corresponding to the literals $A$ and $B$ (that are involved in $F$ and $G$ respectively) simultaneously. The given translation satisfies our intuition. It sends both $x$ and $y$ through channel $z$ followed by the parallel execution of $F$ and $G$.

\paragraph{The $\oplus$ rules} The $\oplus$ operator provides the means to ignore an argument or, consequently, a channel. In the first rule, for example, we expect to receive two names $u$ and $v$ corresponding to the channels for $A$ and $B$ respectively through a common channel $z$. The process ignores the second name $v$ and uses the first one $u$ to send $x$ before invoking $P$. The process for the second rule is symmetric as it ignores the first name $u$ and sends $y$ through $v$. 

\paragraph{The $Cut$ rule} The $Cut$ rule is perhaps the most significant rule as far as process interactions are concerned. We already discussed that applying cut-elimination to the proof corresponds to performing reductions in its $\pi$-calculus translation. Therefore, the $Cut$ rule corresponds to a reduction/interaction between processes $F$ and $G$. Additionally, the interaction will take place through the ports corresponding to the literal being cut, namely $C$. Thus, port $x$ will be connected to port $y$ to form a common channel $z$. The two processes are expected to interact through this common channel $z$ when invoked in parallel. It is worth noting that there are no assumptions made whatsoever about which of the two services will be the receiver and which will be the sender.

Similar informal justifications can be given for rules involving $\parr$ and $\with$, but will be omitted here due to space limitations. We should also note that $\pi$-calculus replication ($!P$) is only used in the translations of the of-course and why-not rules which, as mentioned previously, are not currently part of our system. Despite that fact, our language is sufficiently expressive to describe useful Web Services compositions, as demonstrated in Section \ref{results}.

\subsection{Achieving Web Services composition}
\label{wscomp}

The proofs-as-processes paradigm allows us to create Web Services compositions described using the $\pi$-calculus by performing CLL proofs. In short, the available Web Services descriptions are specified as CLL sentences as we discussed in Section \ref{WSCLL}. Then we construct a CLL description of the requested composite service and attempt to prove it as a conjecture. Using our defined logic, proving the requested service $R$ from the available services $A_i$ corresponds to proving the following goal:
\[
	\infer{R}{A_1 & A_2 & ... & A_n}
\]

Assuming the proof (ie.\ the composition) can be accomplished, the resulting lemma will correspond to a valid logical representation of the requested service. We will have, therefore, proven that such a service exists and can be constructed using the set of available Web Services. Moreover, the $\pi$-calculus translation of the proof will provide a full description of the structure of the composite service.

\section{Results}
\label{results}

We begin the result analysis by explaining the setup of our implementation for the ski example in Section \ref{setup}. This is followed by a brief description of the obtained results in Section \ref{resres} and by a description of the execution of the resulting $\pi$-calculus process as an empirical verification in Section \ref{execution}.

\subsection{Setup}
\label{setup}

Our system is built within the higher order logic (HOL) proof assistant HOL Light. More specifically, we have embedded MALL and the $\pi$-calculus syntax within HOL Light, while making sure proof annotations and process calculus term construction are supported.

The embedding of the $\pi$-calculus is based on the work of Melham in HOL88 \cite{melham1994mechanized}. It includes the basic, polyadic $\pi$-calculus syntax, a few simple functions about names, and substitution. Formalising reductions, structural congruence rules, and bisimulation rules may prove useful for further meta-theoretic reasoning, but is currently beyond the scope of our project.

For the MALL embedding we followed the work of Power et al.\ \cite{power1999working} and Sadrzadeh \cite{sadrzadeh2003modal} in Coq. We follow a similar methodology, although we use multisets instead of lists of sequents, and thus have no need for an \texttt{Exchange} rule to swap the order of sequents in a sentence. 

Supporting $\pi$-calculus proof annotations and enabling process calculus term construction in the style of type theory was a fairly challenging task. We achieved this by including the channel names and the proofs-as-processes translations as logical terms within the embedding of each MALL inference rules. Our custom tactics allow us to accomplish CLL proofs of CLL statements with $\pi$-calculus annotations using these combined rules while constructing the $\pi$-calculus translation simultaneously. Their functionality is based on the use of metavariables as we further explain in the next section.

Having implemented this system as an embedded logic within HOL Light guarantees the correctness of the involved proofs. Given the soundness and correctness proofs of Bellin and Scott for the proofs-as-processes paradigm, this also guarantees the correctness of the composite service, ie.\ that the resulting $\pi$-calculus service will indeed have the expected behaviour. 

We will use the Ski example to demonstrate the functionality of our system. In Section \ref{WSCLL} we presented the CLL translations of the available services as well as the translation of the requested service. Based on the scheme described in Section \ref{wscomp}, achieving the desired services composition is equivalent to proving the requested service as a conjecture. We, therefore, need to prove the following lemma:
\begin{equation}
\label{conjecture}
	\begin{aligned}
	\exists P. \vdash P \cc x \cln PRICE\_LIMIT^\bot,\ y \cln SKILL\_LEVEL^\bot,\ z \cln HEIGHT\_CM^\bot, \\
  w \cln WEIGHT\_KG^\bot,\ t \cln (PRICE\_NOK \oplus EXCEPTION)
	\end{aligned}	  
\end{equation}

We note that the existential quantification of $P$ is not at the embedded level of CLL but rather at the HOL (meta) level. It allows us to find the process $P$ for which this sentence holds, ie.\ the composite service that satisfies this specification.

\subsection{Proof and obtained results}
\label{resres}

The proof of (\ref{conjecture}) is shown in Figure \ref{fig:SProof2}. Note that the CLL propositions are abbreviated, eg.\ $HC$ stands for $HEIGHT\_CM$, and the process calculus annotations have been omitted for a cleaner presentation. Moreover, the \texttt{neg\_eq} step in the proof is an abbreviation of the usage of the $Cut$ rule with a lemma involving negation. As each of the CLL inference rules is applied in the proof, the composite process $P$ is gradually constructed based on the corresponding process calculus translations of the proofs-as-processes paradigm (see Figure \ref{CLLpi}).

\begin{figure}[htbp]
	\centering
	\scriptsize
	\begin{equation}
		\label{proofpart1}
		\infer[\otimes]{\vdash PL^\bot,\ SL^\bot,\ HC^\bot,\ WK^\bot,\ (BR \otimes MO) \otimes LI}{
			\infer[SelMod]{\vdash PL^\bot,\ SL^\bot,\ BR \otimes MO}{}
			&
			\infer[Cut]{\vdash HC^\bot,\ WK^\bot,\ LI}{
				\infer[SelLen]{\vdash HC^\bot,\ WK^\bot,\ LC}{}
				&
				\infer[Cm2Inch]{\vdash LC^\bot,\ LI}{}
			}
		}
	\end{equation}
	\medskip
	\medskip
	\begin{equation}
		\label{proofpart2}
	\infer[Cut\ with\ (\ref{proofpart1})]{\vdash PL^\bot,\ SL^\bot,\ HC^\bot,\ WK^\bot,\ PN \oplus EXE}{
		\infer[neg\_eq]{\vdash((BR \otimes MO) \otimes LI)^\bot,\ PN \oplus EXE}{
			\infer[Cut]{\vdash (BR^\bot \parr MO^\bot) \parr LI^\bot,\ PN \oplus EXE}{
				\infer[\parr]{\vdash (BR^\bot \parr MO^\bot) \parr LI^\bot,\ PU \oplus EXE}{
					\infer[\parr]{\vdash BR^\bot \parr MO^\bot,\ LI^\bot,\ PU \oplus EXE}{
						\infer[SelSki]{\vdash BR^\bot,\ MO^\bot,\ LI^\bot,\ PU \oplus EXE}{}
					}
				}
				&
				\infer[neg\_eq]{\vdash (PU \oplus EXE)^\bot,\ PN \oplus EXE}{
					\infer[\with]{\vdash PU^\bot \with EXE^\bot,\ PN \oplus EXE}{
						\infer[\oplus]{\vdash PU^\bot,\ PN \oplus EXE}{
							\infer[Usd2Nok]{\vdash PU^\bot,\ PN}{}
						}
						&
						\infer[\oplus]{\vdash EXE^\bot,\ PN \oplus EXE}{
							\infer[Id]{\vdash EXE^\bot,\ EXE}{}
						}
					}
				}
			}
		}
	}
	\end{equation}
	\normalsize
	\caption{The proof of the requested service of the Ski example in CLL.}
	\label{fig:SProof2}
\end{figure}

Without delving into too many technical details, we accomplish this by introducing $P$ as a metavariable that is matched to the translation of the first rule being applied to the goal (assuming we are following the proof backwards), ie.\ the $Cut$ rule. This turns $P$ into an instantiation of $\nu z (F_{\vec{u}}[z/x] || G_{\vec{v}}[z/y])$ where any matched variables have also been instantiated. 

Any unmatched variables in this new form are also introduced as metavariables that are in turn instantiated in the next proof steps. For example, $G$ eventually becomes the process corresponding to part (\ref{proofpart1}) of the proof (see Figure \ref{fig:SProof2}). It is first introduced as a new metavariable that is gradually instantiated as the proof progresses. 

If one of the available services' CLL statements (see Figure \ref{fig:AvailServCLL}) is used to match one of the CLL sentences at the top of the proof tree, this will instantiate one of the metavariables and the service will then be introduced as a component in the $\pi$-calculus representation of the composite service.

Any unmatched metavariables at the end of the proof are left as fresh, free variables in the $\pi$-calculus result. In the above example of $\nu z (F_{\vec{u}}[z/x] || G_{\vec{v}}[z/y])$, $z$ is the variable representing the channel connecting $F$ and $G$ and will never be matched as it never appears in the proof. It is in fact kept as a fresh variable and renamed to $z_1$ to avoid variable clashes in our result.

The $\pi$-calculus term that was constructed following this methodology by our proof for the Ski example (denoted as $Composition(smp,sms,slh,slw,t,puc,exc)$) is presented in Figure \ref{fig:SkiPiResult}. It is immediately apparent that the complexity of this term prohibits any attempt to fully analyse its functionality on paper. Only some of its parts are clear and can be analysed. For example, the subterm $(\nu\ z_3)(SelLen[z_3/sll]\ ||\ Cm2Inch[z_3/cic])$ represents the interaction between port $sll$ of the $SelLen$ service and port $cic$ of the $Cm2Inch$ service\footnote{Note how both $sll$ and $cic$ are substituted by $z3$ in order to accomplish this interaction.}. Essentially, this is a composite subprocess that selects the ski length in inches (rather than in centimeters). It resulted from the application of the $Cut$ rule using lemma (\ref{proofpart1}) in the proof.
\begin{figure}[tbp]
\centering
\begin{minipage}{.65\linewidth}
\begin{flushleft}
\hspace*{0mm}  $Composition(smp,sms,slh,slw,t,puc,exc)\ \equiv$ \\
\hspace*{5mm}  $(\nu\ z_1)$ \\
\hspace*{10mm}  $  ((\nu\ smo,cii)$ \\
\hspace*{15mm} $    (\overline{z_1}\langle smo,cii\rangle.$ \\
\hspace*{15mm} $    (SelMod\ ||\ (\nu\ z_3)(SelLen[z_3/sll]\ ||\ Cm2Inch[z_3/cic])) $\\
\hspace*{10mm}  $  )\ ||\ (\nu\ z_4) $\\
\hspace*{15mm} $    (((z_1(x_5,ssl).x_5(ssb,ssm).SelSki[z_4/sso])\ ||\   $\\
\hspace*{15mm} $    (\nu\ u_7,v_7)$\\
\hspace*{20mm} $      ((\overline{z_4}\langle (u_7,v_7)\rangle. $\\
\hspace*{25mm} $        ((u_7(unu).(\nu\ unn)(t(u_8,v_8).\overline{u_8}\langle unn \rangle.Usd2Nok)) + $\\
\hspace*{30mm} $    (v_7(y_7).(\nu\ y_9)(t(u_9,v_9).\overline{v_9}\langle y_9\rangle.y_7(a_{10}).\overline{y_9}\langle a_{10}).0))) $\\
\hspace*{15mm} $))$ \\
\hspace*{10mm}  $)$\\
\hspace*{5mm}  $)$\\
  \end{flushleft}
  \end{minipage}
	\caption{The resulting $\pi$-calculus formula from the Ski proof.}
	\label{fig:SkiPiResult}
\end{figure}

It is worth noting that the resulting services composition makes no assumptions about the form of the component services. In our example from Figure \ref{fig:SkiPiResult}, the $\pi$-calculus term includes component processes such as $SelLen$, $SelMod$, $SelSki$, etc.\ as ``black boxes'' with hidden functionality. The only known properties of the component services are the input and output ports as defined in their CLL representation. For example, $SelLen$ is defined as follows (see Figure \ref{fig:AvailServCLL}):
\[
\vdash SelLen \cc slh \cln HC^\bot,\ slw \cln WK^\bot,\ sll \cln LC
\]
Therefore, it is only assumed that it has input ports $slh$ and $slw$ and output port $sll$.

\subsection{Execution}
\label{execution}

Once the $\pi$-calculus composition is extracted, the next step is to execute the composed service. This can be accomplished by translating the $\pi$-calculus representation in a more commonly used, executable Web Services description language such as the previously mentioned WSDL, BPEL4WS, or OWL-S. 

In our project so far, before undertaking the formal translation of $\pi$-calculus to any other model, we are focusing on visualising and checking the results empirically. This is accomplished by introducing concrete $\pi$-calculus representations for each of the available services and simulating their execution by invoking the $\pi$-calculus reductions.

We have, therefore, constructed a set of mappings for a systematic interpretation of CLL judgements to $\pi$-calculus processes. These mappings provide an intuitive interpretation that satisfies the corresponding properties and follows the expected behaviour (as far as $\pi$-calculus reductions are concerned). We present them in Figure \ref{proctrans}.

\begin{figure}[htbp]
	\centering
		\begin{tabular}{rl}
			$A$ & $\overline{a}\langle ...\rangle .0$ \\
			$A^\bot$ & $a(...).0$ \\
			$A \otimes B$ & $(\nu a,b)(\overline{z}\langle a,b \rangle.(\overline{a}\langle ...\rangle.0\ ||\ \overline{b}\langle ...\rangle .0))$ \\
			$A^\bot \parr B^\bot$ & $z(a,b).(a(...).0\ ||\ b(...).0)$\\
			$A \oplus B$ & $(\nu a,b) (z(u,v).(\overline{u}\langle x \rangle.\overline{x}\langle ... \rangle.0 + \overline{v}\langle y \rangle.\overline{y}\langle ... \rangle.0))$ \\
			$A^\bot \with B^\bot$ & $(\nu a,b) (\overline{z}\langle u,v \rangle.(u(x).x(...).0 + v(y).y(...).0))$\\
		\end{tabular}
			\caption{Translations of CLL terms to $\pi$-calculus.}
			\label{proctrans}
\end{figure}

Following these empirical translations, we introduce $\pi$-calculus terms for each of the available services. We, therefore, instantiate the ``black boxes'' in our initial result with an executable $\pi$-calculus term. Moreover, we introduce a $\pi$-calculus term for a $Request$ service. This service can be viewed as the client for the requested composite service. It will interact with the $\pi$-calculus term of the latter to verify its functionality. In this particular example, it will provide the expected input, ie.\ the price limit, skill level, height and weight, and expect the desired output, namely the price in NOK or an exception. We present the introduced $\pi$-calculus services in Figure \ref{fig:AvailServPi}. The parallel composition of the $Request$ service and the derived composite service $Composition$ from Figure \ref{fig:SkiPiResult} is introduced as the $Main$ service to complete our model.

\begin{figure}[htbp]
\centering
\begin{minipage}{.95\linewidth}
\begin{flushleft}
\footnotesize
$SelLen (slh,slw,sll,lc) \equiv slh(hc).slw(wk).\overline{sll}\langle lc \rangle.0$ \\
	\medskip
$Cm2In (cic,cli,li) \equiv cic(lc).\overline{cli}\langle li\rangle .0$ \\
	\medskip
$Usd2Nok (unu,unn,pn) \equiv unu(pu).\overline{unn}\langle pn\rangle .0$ \\
	\medskip
$SelMod (smp,sms,smm,br,mo) \equiv smp(pl).sms(sl).(\nu\ smb,smo)(\overline{smm}\langle smb,smo\rangle .\overline{smb}\langle br\rangle .\overline{smo}\langle mo\rangle .0)$ \\
	\medskip
$SelSki (ssb,ssm,ssl,sso,pu,ex) \equiv (\nu\ ssp,sse)(ssb(br).ssm(mo).ssl(li).sso(u,v).(\overline{u}\langle ssp\rangle .\overline{ssp}\langle pu\rangle .0 + \overline{v}\langle sse\rangle .\overline{sse}\langle ex\rangle .0))$ \\
	\medskip
$Request(smp,pl,sms,sl,slh,hc,slw,wk,t,puc,exc) \equiv$ \\
\hspace*{1cm}  $\overline{smp}\langle pl\rangle .\overline{sms}\langle sl\rangle .\overline{slh}\langle hc\rangle .\overline{slw}\langle wk\rangle .\overline{t}\langle puc,exc\rangle .(puc(x).x(pu).0 + exc(y).y(ex).0)$ \\
	\medskip
$Main () \equiv Request(smp,pl,sms,sl,slh,hc,slw,wk,t,puc,exc)\ ||\ Composition(smp,sms,slh,slw,t,puc,exc)$
		\end{flushleft}
		\end{minipage}
		\normalsize
	\caption{The available services and the $Request$ service for the Ski example defined as $\pi$-calculus processes.}
	\label{fig:AvailServPi}
\end{figure}

It is important to note that these particular concrete representations are not unique in $\pi$-calculus. However, they were designed to be as simple and straightforward as possible. Our aim is to empirically confirm the correctness of the constructed composition and its associated information flow. Since the composition makes no assumptions about the involved component services (apart from the names of the channels through which the services can be interfaced), any valid $\pi$-calculus representation of these services is acceptable. 

\begin{figure}[htbp]
\centering
	\includegraphics[scale=.3]{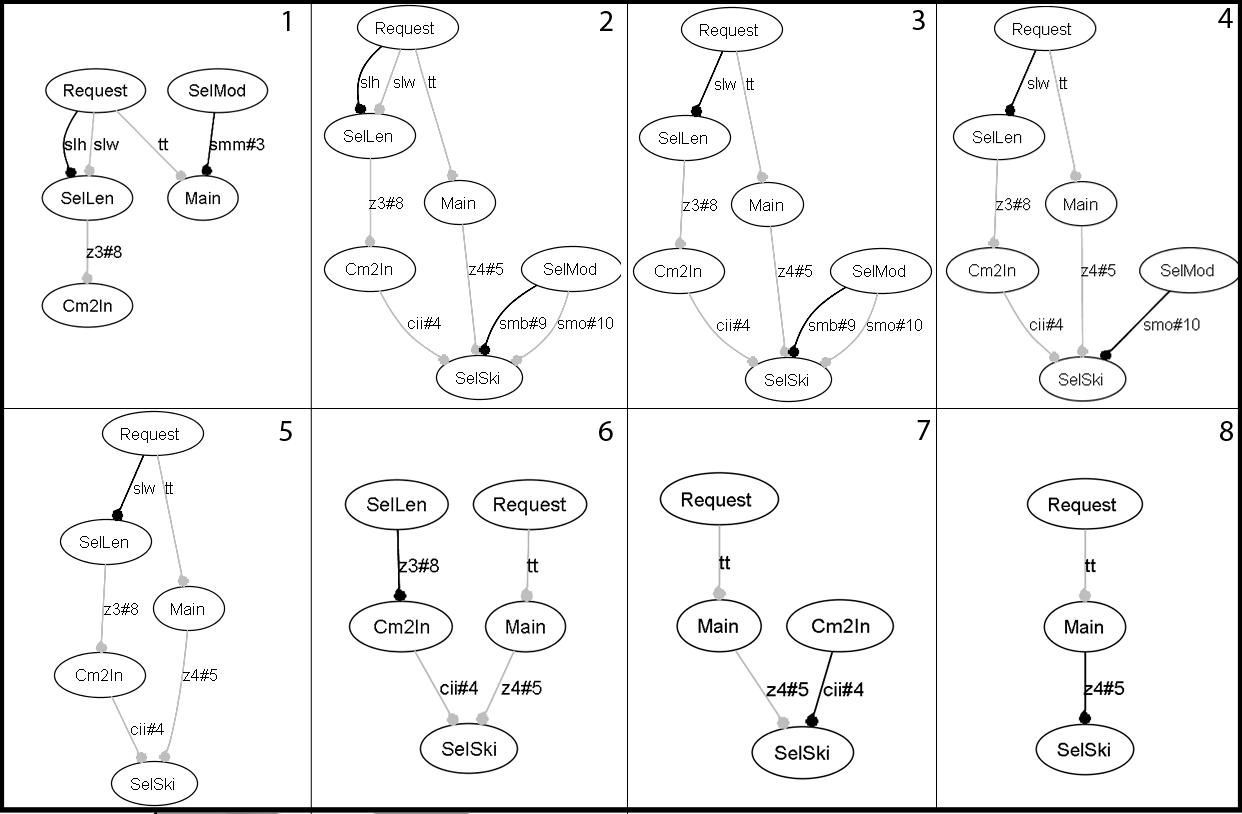}
	\caption{Consecutive snapshots of the Ski example $\pi$-calculus result taken from PiVizTool. }
	\label{fig:PiViz}
\end{figure}

Our complete set of services was given as input to the PiVizTool (see Section \ref{pidesc}). This system allows the visualisation of the various services and their interactions (corresponding to $\pi$-calculus reductions). Eight consecutive snapshots (out of a total of 17) from the resulting visualisation are shown in Figure \ref{fig:PiViz}. Each edge represents a possible interaction between two agents. The grey edges represent interactions that are currently blocked whereas the black edges represent interactions that can occur immediately on the next execution step. Each snapshot is the result of applying one $\pi$-calculus reduction in the previous state.

For example, in snapshot 6 of Figure \ref{fig:PiViz}, the $SelLen$ service interacts with $Cm2In$ through channel $z3$ (automatically renamed to $z3\#8$ by PiVizTool to avoid clashes). Essentially, this corresponds to the conversion of the output of the $SelLen$ service from centimeters to inches via the $Cm2In$ service as we discussed in the previous section. The result after the interaction is shown in snapshot 7.

The PiVizTool simulation plays an important role towards empirically verifying that our result satisfies the requested service without the initial need for a more concrete execution model. Our process behaves as expected indicating that the available services have been succesfully composed.

Additionally, PiVizTool provides the definition and execution trees of the composed service. These are tree visualisations of the definitions and execution sequence for each of available services respectively. These trees are also helpful towards understanding and analysing the behaviour of the composite service.

\section{Related work}
\label{relatedwork}

There are two main directions in the research over Web Services composition: one involves the use of workflow techniques while the other relies on AI planning \cite{hendler1990ai}. 

The workflow techniques rely on the requester building an abstract process model, including a set of required tasks and their data dependencies. The aim is then to build a graph of atomic services that can fulfill this role. An example of such a workflow-based system is EFlow \cite{casati2000eflow}. 

In the AI planning approach, services are considered as actions with specified preconditions and effects. A planner then attempts to discover the appropriate combination of actions that will lead to a goal state starting from an initial state. An example of such system is SWORD \cite{ponnekanti2002sword} which uses rule-based planning for the composition of services.

Theorem proving techniques, such as the one used in the current work, are considered part of the planning approach to Web Services composition. There have been multiple attempts at using theorem provers in this context. Waldinger, for instance, based his work on automated deduction and program synthesis \cite{waldinger2001wac}. He used the theorem prover SNARK \cite{stickel2000guide} to provide proofs for service composition problems described in classical first-order logic. Lammermann worked on structural synthesis of program, a deductive approach that utilises intuitionistic propositional logic \cite{lammermann2002rsc}. Finally, Rao et al.\ used propositional Linear Logic theorem proving with DAML-S based proofs \cite{rao-semantic,rao2006composition}. 

The latter has been the main motivation for this work. However, after carefully analysing the work of Rao et al., we detected a number of potential inconsistencies. For example, the process calculus being used is an extension of the $\pi$-calculus. However, no guarantees are given that the two calculi are equivalent or that the Bellin and Scott proofs are valid for the extended process calculus. Additionally, they use Intuitionistic Linear Logic in a two-sided sequent calculus, which is also not guaranteed to be equivalent to the one-sided CLL approach of Bellin and Scott. Moreover, a number of so-called ``structural congruence rules'' that contain both CLL terms and channel names were introduced. These rules were not formally derived and their syntax can easily result in an incorrect interpretation. Finally, even though the system theoretically supports non-functional properties and exponentials, the proof of the Ski example ignores them. Non-functional properties are crucial towards a quality-driven composition, whereas adding the exponentials would make the logic undecidable. 

Using our system and the higher order logic backround of HOL Light we were able to provide a rational reconstruction of Rao's work. This includes a formal interpretation (using CLL and the $\pi$-calculus) of some of Rao's introduced concepts such as composite and optional ports and channels and the verification of some their properties and structural congruence rules. The lack of published code and more examples made this a fairly challenging undertaking.

\section{Conclusion and Future work}
\label{conclusion}

We have described our efforts towards the implementation of a rigorous framework for Web services composition using the higher order logic proof assistant HOL Light. Our approach is based on the proofs-as-processes paradigm originally introduced by Abramsky, Bellin and Scott.

In contrast to the work of Rao et al., we have attempted to remain faithful to the original theory of Bellin and Scott by using CLL in conjunction with the standard polyadic $\pi$-calculus syntax. Our implementation has shown some promising results and there is sufficient room for improvements and further work. In particular, interesting properties such as liveness, safety, and deadlock-freedom have not been investigated in this work yet. These, along with the automation of our CLL proofs and further evaluation of the system, are part of our next goals.

The Bellin and Scott translation has the interesting property of keeping the $\pi$-calculus annotations and the CLL proof completely independent of each other. This means that the CLL proof is not affected by the attached process calculus terms. In fact, if we completely remove the annotations, the proofs are perfectly valid CLL proofs. We plan to exploit this property in our attempt to utilise external tools to automate our proofs. Our effort will focus on finding and using tools (such as llprover \cite{tamura1998user}) that can perform automated CLL proofs and return the entire proof script, which can then be integrated and verified by our HOL Light framework.

We will also be focusing on further system evaluation. Our implementation test set is currently using Rao's Ski example as its main case. In the next stages of this work we will gather practical examples from existing web services projects, such as the \textsc{Sensoria} Project \cite{sensoria2006}, where verified composition is desirable. Moreover, we will compare our system with related systems in the field. We also note that it is important to find examples of composable web services with non-functional properties. The potential of our framework to incorporate such properties in the composition sets it apart from related work. 

Another important part of our future work is to establish formal translations from the composite services produced by our system in $\pi$-calculus to more widely used and more concrete Web Services description languages, including WSDL and OWL-S. This involves the proper translation of the composite service's control flow, including notions such as sequence that are not explicitly represented in $\pi$-calculus.

In conclusion, we believe that our work contributes to both the Web Services and theorem proving/formal verification research areas. On the one hand, we are working towards a fully verified Web services composer using theorem proving techniques while promoting theoretical research in this area. On the other hand, the variety of tools and tactics that we have developed may provide a reusable and extensible library, which may prove useful in other formal verification or theorem proving projects in the area and beyond.

\section{Acknowledgement}

We would like to thank the anonymous reviewers for their helpful and constructive comments.

\bibliographystyle{eptcs}
\bibliography{WWV}

\end{document}